\documentclass{elsart}
\usepackage[dvips]{graphicx}

\newcommand{\bbb}{Bi$_{2}$Sr$_{2}$Ca$_{2}$Cu$_{3}$O$_{10}$}

\newcommand{\hTc}{high-$T_c$}
\usepackage{amssymb}
\begin{document}
\begin{frontmatter}
\title{Scaling properties of the optical conductivity of Bi-based cuprates}
\author{D. van der Marel\corauthref{cor1}},
\corauth[cor1]{Corresponding author}
\ead{dirk.vandermarel@physics.unige.ch}
\ead[url]{http://optics.unige.ch}
\author{F. Carbone, A. B. Kuzmenko, E. Giannini}
\address{D\'epartement de Physique de la Mati\`ere Condens\'ee, Universit\'e de Gen\`eve,
Quai Ernerst-Ansermet 24, Gen\`eve 4, Suisse}
\runtitle{optical conductivity scaling}
\runauthor{D. van der Marel}
\begin{abstract}
We present novel infrared optical conductivity data on the three
layer high T$_c$ superconductor Bi$_2$Sr$_2$Ca$_2$Cu$_3$O$_{10}$
at optimal doping. We extend the analysis of an earlier
publication, providing a universal scaling function
$\sigma(\omega,T)=T^{-1}g(\omega/T)$ for the optical conductivity.
In the present manuscript we obtain a good scaling collapse of the
experimental curves on the $g(\omega/T)$ over a wide range of
values of $\omega/T$ (at least in range 0 to 10), if we assume
that $g(\omega/T)$ is superimposed on a non-universal background
which is temperature independent. We obtain the same result, if in
our analysis we allow this background to have a $T^2$ temperature
dependent correction. The most striking property of $g(\omega/T)$
is, that it corresponds to a scattering rate which varies linearly
as a function of temperature, but which is independent of the
frequency.
\end{abstract}
\end{frontmatter}
\section{INTRODUCTION}\label{intro}
The temperature and frequency dependence of the free carrier
optical conductivity of the \hTc\ cuprates has been the subject of
debate since the discovery of the high temperature
superconductors. The interpretation of these data is relevant and
important, because the departure from simple Drude behavior
reflects the nature of the interactions responsible for the
transport properties of the normal state and of the
superconductivity occurring in these materials. Recently some of
us have pointed out two peculiar aspects of the infrared optical
conductivity in the normal state\cite{marel03}:
\begin{itemize}
\item In the far infrared range the optical conductivity of
optimally doped cuprates is characterized by a universal
$\omega/T$ scaling function of the form
\begin{equation}\label{scaling}
  \sigma(\omega) = T^{-1} g(\omega/T)
\end{equation}
where the function $g(x)$ is to very good approximation given by
$g(x)=C/(1-iAx)$ with $A\approx 0.8$ for $x=\omega/T<1.5$. This
corresponds to a Drude response where the scattering rate has a
linear temperature dependence. A consequence of this behavior is a
collapse of the spectra when $T\sigma_1(\omega,T)$ is plotted as a
function of $\omega/T$ for different temperatures.
\item At $\omega/T\approx 1.5$ a cross-over appears to take place
to a power law behavior, previously pointed out in Refs.
\cite{schlesinger90,elazrak94,baraduc96,anderson97}. For
$\omega/T>1.5$ a collapse of the spectra plotted versus $\omega/T$
required that the conductivity is multiplied no longer with the
temperature $T$, but with either $T^{0.5}$ or $\omega^{0.5}$. This
change of behavior is also evident from a plot of the phase of the
optical conductivity ($\arctan(\sigma_2/\sigma_1)$), which
displays a plateau with a phase angle of approximately 60 degrees
for all temperatures and frequencies in access of $k_BT$. In the
same frequency range the absolute value of the conductivity,
$¦\sigma(\omega)¦$, follows a power law
behavior,$|\sigma(\omega)|\propto \omega^{-2/3}$. Both the
frequency dependence of $¦\sigma(\omega)¦$ and of the phase angle
are manifestations of the fact that the optical conductivity
follows approximately a power law for frequencies in the range
$T<\omega<0.7eV$. \end{itemize}
The exponent $\alpha = 2/3$ is rather different from 0.5, as given
by the 'cold spot' model\cite{ioffe98} and closer to Anderson's
result based on the concept of spin-charge
separation\cite{anderson97}. Experimentally the $\omega/T$ scaling
and constant phase angle are most closely obeyed for samples close
to optimal doping. As such they appear to be a direct
manifestation of quantum critical behavior when the doping is
tuned to match exactly a quantum phase transition\cite{sachdev99}.
However, it remained unclear whether a single scaling function
could be defined covering both aspects 1) and 2) of the optical
conductivity.

Here we extend the analysis of Ref. \cite{marel03} by introducing
a novel transformation, allowing us to obtain the $\omega/T$
scaling function for all frequencies. The relations for this
transformation rely on the derivative with respect to temperature
of the real and imaginary part of the dielectric function. For
such a procedure it is important to have a dense sampling of
temperature, excellent signal-to-noise ratio, and a very small
spurious time-dependent drift of the data, in particular in the
frequency range below 1 eV. Due to recent improvements of our
infrared spectrometer, in particular in the range of 0.5 to 0.8
eV, the aforementioned requirements can be met. The best result
obtained in our group in this respect is a novel data
set\cite{carbone06a} on single crystals of tri-layer \bbb grown at
the University of Geneva\cite{giannini04}. In this manuscript we
will therefore apply the novel transformation to the data of
Ref.\cite{carbone06a}.

\section{Novel transformation to obtain $\omega/T$ scaling function}
The basic assumption which we will investigate is whether the
optical conductivity representing the free charge carriers follows a
universal scaling relation of the form\cite{phillips05}:
\begin{eqnarray}
\sigma(\omega,T)= \frac{1}{T^{\nu}}g\left(\frac{\omega}{T}\right)
\mbox{  ;  }\nu \equiv \frac{z}{2-d}
\end{eqnarray}
where $d$ is the relevant dimension and $z$ the dynamical exponent
\cite{sachdev99}. As pointed out by Philips and Chamon, T-linear
resistivity requires the unphysical assumption that the dynamical
exponent $z$ is negative. Consequently, no consistent account of
T-linearity is possible if the quantum critical modes carry the
electrical charge. Apparently, either the T-linear resistivity is
not directly linked to a quantum phase transition, or "quantum
critical scenarios must relinquish the simple single scale
hypothesis to explain the resistivity law in the
cuprates."\cite{phillips05} Despite this deficiency we will treat
$\nu$ as an adjustable scaling parameter, which can have the value
$\nu=1$. It is clear that it is no longer justified in this case
to assume that $\nu = z/(2-d)$.

Before continuing we wish to make one general observation about
the part of the optical conductivity described by $g(\omega/T)$:
The integrated optical spectral weight is proportional to
$T^{1-\nu}$. We already know that in the case of interest for this
paper $\nu=1$, because at optimal doping the resistivity,
$\rho\propto T^{\nu}$, is a linear function of the temperature.
Taken together $\omega/T$-scaling and T-linear resistivity imply
that the integrated spectral weight presented by $g(\omega/T)$ has
a constant value for all temperatures. Experimentally this has to
be confirmed by observing the collapse of $T\sigma(\omega,T)$
plotted as a function of $\omega/T$. In section \ref{results} we
will see that upon close inspection, a few percent spectral weight
variation between T$_c$ and 300 K is not excluded. The optical
conductivity integrated between $0$ and $1.25$ eV is known to
exhibit a temperature dependence proportional to $1-bT^2$, with $b
\approx 5\cdot 10^{-7} K^{-2}$\cite{molegraaf02}. In order to
include the possibility that the integrated free carrier weight
has a correction proportional to $T^2$, we have to multiply
$g(\omega/T)$ with a factor $1-bT^2$.

\begin{figure}[ht]
   \centerline{\includegraphics[width=8cm,clip=true]{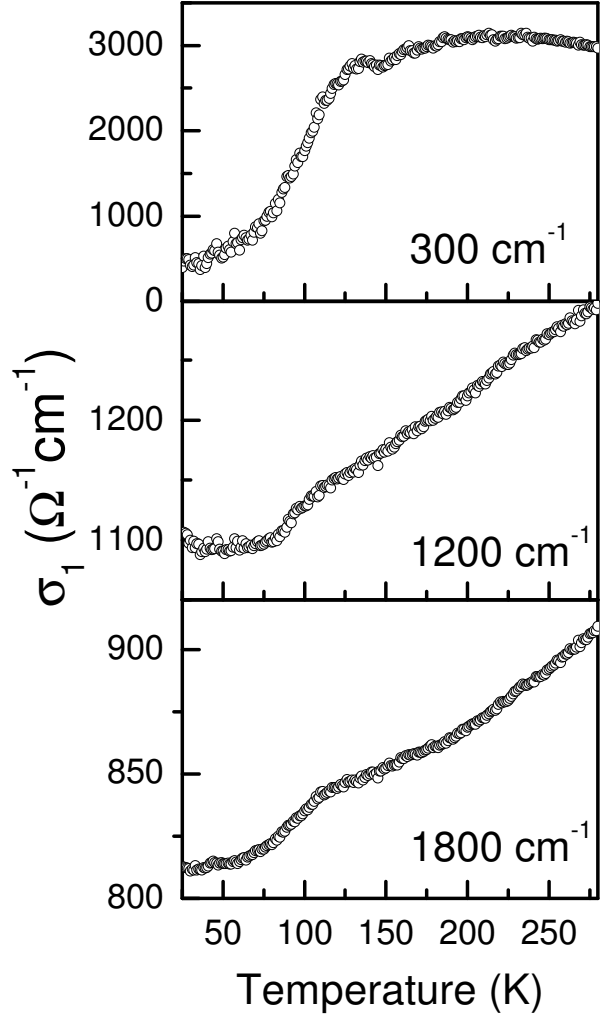}}
   \caption{Temperature dependence of the ab-plane optical conductivity of Bi2223 for three selected frequencies.}
   \label{sigmaT}
\end{figure}
According to Ref. \cite{marel03} the function $g(\omega/T)$ obeys
a Drude form, at least for $\omega/T < 1.5$, but here we are
interested to find out more about the continuation beyond this
range of $\omega/T$. One possible obstacle is, that at higher
frequencies the scaling behavior becomes overshadowed by 'regular'
({\em i.e.} unrelated to quantum critical behaviour) contributions
to the optical conductivity, notably in the range above about 1 eV
where interband transitions dominate the optical conductivity.
Because interband transitions present a fixed energy scale, for
those high frequencies there is no shadow of doubt that the
temperature is not the only relevant scale in the optical
conductivity. For frequencies below 1 eV, but still above $k_BT$
the situation is less clear: Interband transitions are negligible
in this range, but the optical conductivity in this range strongly
exceeds the values obtained by extrapolating the Drude behaviour
seen in the far-infrared range. The observed $\sigma(\omega)$
below 1 eV therefore appears to be an intrinsic part of the free
carrier response. Yet, it needs to be established from the
experimental data whether this part is described by the same
$g(\omega/T)$ function which is found for $\omega/T < 1.5$.
Experimentally\cite{molegraaf02,toschi05} the optical conductivity
in the mid-infrared and visible part of the spectrum of optimally
doped samples has a temperature dependence of the form
$\sigma(\omega,T)=\sigma^{(0)}(\omega)+\sigma^{(2)}(\omega)T^2$.
In Fig. \ref{sigmaT} of the temperature dependence of
$\sigma_1(\omega)=\omega/(4\pi)\mbox{Im}\epsilon(\omega)$ and
Re$\epsilon(\omega)$ are shown. The non-monotonous temperature
dependence in the normal state for 300 cm$^{-1}$ is an immediate
consequence of the fact that in this range the optical
conductivity is dominated by a Drude peak the width of which
varies linearly as a function of temperature. On the other hand
temperature dependence is observed in the optical conductivity at
least up to $20000$ cm$^{-1}$
\cite{molegraaf02,carbone06a,santander03,santander04,deutscher05,kuzmenko05a},
which vastly exceeds the range of the Drude peak. Moreover the
observed temperature dependence is incompatible with the narrowing
of the Drude peak: For $\omega\gg T$ the Drude function
$\sigma_1\propto T/(\omega^2+T^2)$ crosses over to a T-linear
temperature dependence. Indeed this is seen for example at 1200
cm$^{-1}$ (see Fig. \ref{sigmaT}), but at higher frequencies the
temperature variation becomes dominated by a $T^2$ term. Although
the $T^2$ correction remains small relative to $\sigma_1$ in all
cases, its presence in the experimental data motivates us to
explore the following decomposition of the optical conductivity
\begin{eqnarray}
 \sigma(\omega,T)&=&
 \frac{1-bT^2}{T^{\nu}}g\left(\frac{\omega}{T}\right) +
 \sigma^{(0)}(\omega)+\sigma^{(2)}(\omega)T^2
 \label{decomposition}
\end{eqnarray}
\subsection{Temperature independent regular
conductivity}\label{stage1}
Because the $T^2$ corrections $b$ and $\sigma^{(2)}(\omega)$ are
small, we will first work out the consequences in the limit where
they are zero. Since here we are interested in the behavior near
optimal doping, where the resistivity is T-linear, we consider the
case where $\nu=1$. Then Eq.\ref{decomposition} reduces to
\begin{eqnarray}
\sigma(\omega,T)=T^{-1}g\left(\omega/T\right)+\sigma^{(0)}(\omega)
\end{eqnarray}
In order to eliminate the temperature independent term
$\sigma^{(0)}(\omega)$, we start by taking the temperature
derivative of this expression. For the evaluation of $dg/d\omega$
and $dg/dT$ we use the fact that the function $g$ depends only on
a single variable $x$, defined as the ratio $x=\omega/T$.
\begin{eqnarray}
 \frac{\partial\sigma(\omega,T)}{\partial T}&=&
 -\frac{1}{T^2}\left\{g+\frac{\omega}{T}g'\right\}=-\frac{1}{T^2}\frac{d}{dx}(xg)
 \label{dec1}
 \end{eqnarray}
Because $xg$ is a function the $\omega$ and $T$ dependence of
which enters only as the ratio $x=\omega/T$, we are allowed to
substitute $\omega$ for $x$ in taking the derivative. Thus, using
the chain-rule: $d(xg)/dx = -T^{-1}d(xg)/d\omega$. Inserting this
for the righthand side of Eq.\ref{dec1} we obtain
$-Td\sigma/dT=d(xg)/d\omega$. The integration of both sides of
this equation results in an expression which relates $xg(x)$ on
one side of the expression to the frequency-integral of
$d\sigma(\omega)/dT$ on the other side. Multiplying both sides
with $1/x=T/\omega$ finally gives the following transformation of
the complex optical conductivity function
\begin{eqnarray}
 g(\omega,T)=-\frac{T^2}{\omega}\int_o^{\omega}\frac{d\sigma(\omega',T)}{dT}d\omega'
 \label{g1}
 \end{eqnarray}
In the above an important starting assumption was, that
$g(\omega,T)$ depends on the ratio $\omega/T$. It may therefore
look somewhat strange that in Eq.\ref{g1} we write it as a
function of frequency and temperature. However, we want to feed
the experimentally measured optical conductivity into the
righthand side of the above transformation, providing $g$ on
output. {\em A priori} there is no guarantee that the experimental
optical conductivity is of the form assumed in Eq.
\ref{decomposition}. In fact this is the model we like to test by
plotting the output of the transformation, $g(\omega,T)$, as a
function of $\omega/T$ for different temperatures.
%
%
%
%
%
%
%
\subsection{Temperature dependent spectral weight of the scaling
function}\label{stage3}

If $\sigma^{(2)}(\omega)$ is different from zero in
Eq.\ref{decomposition} and $b\neq0$, we have no analytical
expression for the transformation. However, it is possible, to
perform a least-square fit of the reflectivity and ellipsometry
data for all temperatures and frequencies at once, inserting in
Eq.\ref{decomposition} a multi-oscillator Kramers-Kronig
consistent composition \cite{kuzmenko05} for the three terms
$g\left(\omega/T\right)$, $\sigma^{(0)}(\omega)$ and
$\sigma^{(2)}(\omega)$.

\section{Experimental procedures}

\subsection{Experiments}\label{SectionExperiment}

Bi$_2$Sr$_2$Ca$_2$Cu$_3$O$_{10}$ (Bi2223) crystals were grown by the
Travelling Solvent Floating Zone method (TSFZ) in an Image Furnace
as described in Refs. \cite{giannini04,clayton04}. All optical
measurements of Bi2223 were performed on an ab-oriented single
crystal at temperatures from 4 to 300 K. In the frequency range 6000
- 37000 cm$^{-1}$ (0.75 - 4.6 eV) we used ellipsometry, which yields
directly the dielectric constant and the optical conductivity. Since
the ellipsometric measurement is done at a finite angle of incidence
(in our case 74$^\circ$), the admixture of the c-axis component of
the dielectric tensor has to be accounted. We used our own c-axis
dielectric function of Bi-2223 to convert the measured
'pseudo-dielectric' function to the 'true' ab-plane dielectric
function using a proper mathematical procedure. Since the
pseudo-dielectric function is rather insensitive to the c-axis
dielectric constant in this range, the small temperature changes of
the latter do not introduce significant errors to the obtained
temperature-dependent in-plane spectra.

In the far-infrared and mid-infrared ranges (100 - 6000 cm$^{-1}$,
or 12 - 750 meV) the reflectivity at a near-normal angle of
incidence was measured. In order to obtain the optical
conductivity in this region, we used a variational routine
\cite{kuzmenko05}: a causal ({\em i.e.} Kramers-Kronig-consistent)
dielectric function, which gives the best detailed match to the
experimental reflectivity at low frequencies and {\em
simultaneously} to the complex dielectric function at higher
energies. This procedure is more accurate than the conventional
Kramers-Kronig reflectivity transformation, since the knowledge of
the two independent quantities at higher energies (instead of one)
allows one to better 'anchor' the unknown phase of the
reflectivity at low energies. The routine also used the soft X-ray
reflectivity spectra of Bi2212 in order to approximately set the
high-frequency conductivity peaks.

\section{Results}\label{results}
\begin{figure}[ht]
   \centerline{\includegraphics[width=8cm,clip=true]{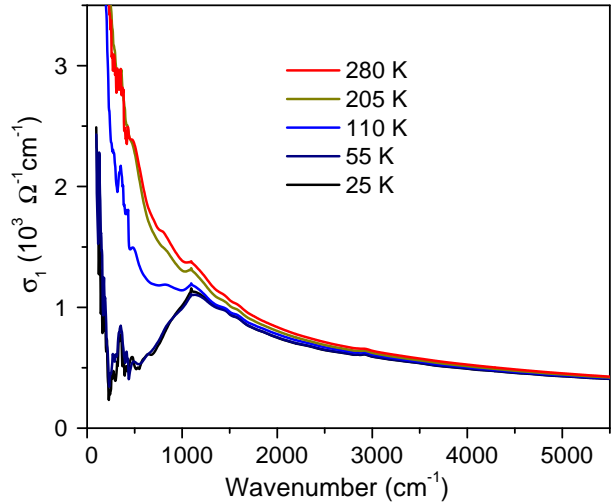}}
   \caption{Optical conductivity spectra along the ab-plane of optimally doped Bi2223 at selected temperatures.}
   \label{FigSig2223}
 \end{figure}
\begin{figure}[ht]
   \centerline{\includegraphics[width=8cm,clip=true]{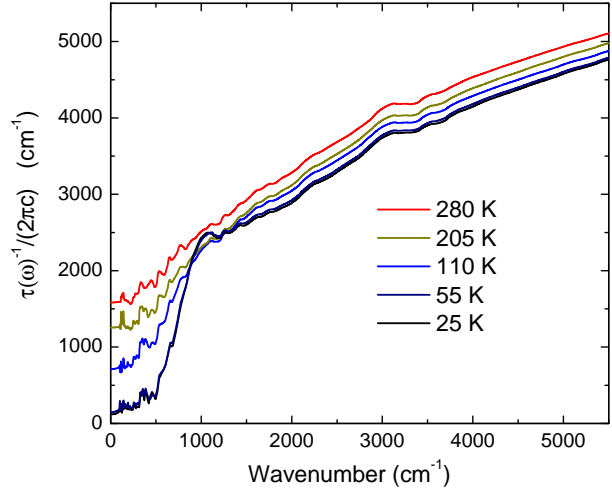}}
   \caption{In-plane frequency-dependent scattering rate of Bi2223.}
   \label{FigTau}
\end{figure}
\begin{figure}[ht]
   \centerline{\includegraphics[width=8cm,clip=true,angle=0]{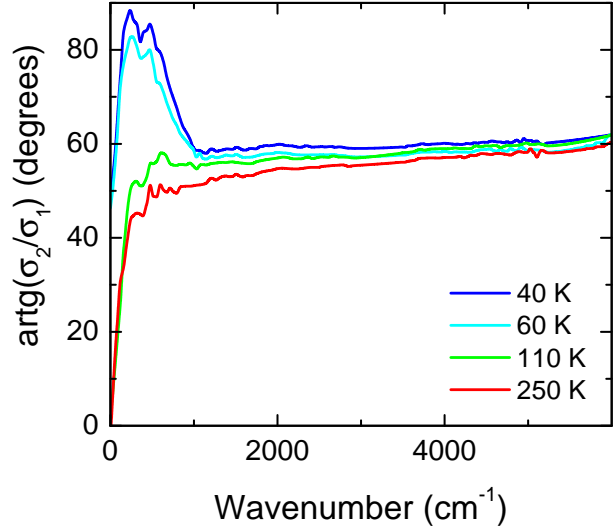}}
   \caption{Phase of the in-plane optical conductivity of Bi2223.}
   \label{FigPhase}
\end{figure}
In Fig. \ref{FigSig2223} we present the in-plane optical
conductivity spectra of Bi2223. Using the extended Drude formalism,
we can obtain the frequency-dependent scattering rate:
\begin{equation}\label{tau}
  \tau^{-1}(\omega) =
  \frac{\omega_p^2}{4\pi}\mbox{Re}\left[\frac{1}{\sigma(\omega)}\right]=\frac{\omega_p^2}{\omega}\mbox{Im}\left[\frac{1}{\epsilon_{\infty}-\epsilon(\omega)}\right]
\end{equation} \noindent
%
%
%
The plasma frequency $\omega_{p}$ was taken to be 20600 cm$^{-1}$,
which gives $m^{\ast}(\omega)/m\equiv
\omega_p^2\omega^{-2}Re\{\epsilon_{\infty}-\epsilon(\omega)\}^{-1}\approx
1$ at 1 eV. The value for $\epsilon_{\infty}=4.5$ was obtained
from a Drude-Lorentz fit to the interband transitions above 1 eV.
The result, shown in Fig.\ref{FigTau}, indicates that the
scattering rate has a power law type frequency dependence
$\tau^{-1}\propto\omega^{\eta}$ with $\eta\approx 2/3$ as for
optimally doped Bi2212. The phase of the optical conductivity,
shown for a few temperatures in Fig. \ref{FigPhase} is close to 60
degrees and almost constant, which is also similar to the phase of
Bi2212 at optimal doping.\cite{marel03}.

\begin{figure}[ht]
   \centerline{\includegraphics[width=8cm,clip=true,angle=0]{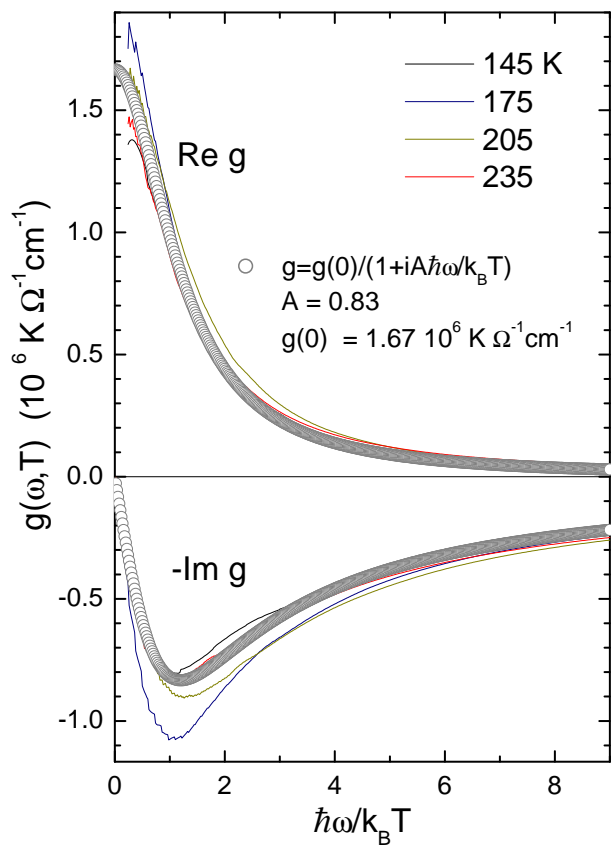}}
   \caption{Real and imaginary parts of the scaling function $g(\omega,T)$ of Bi2223,
   calculated for 4 different temperatures from the experimental data using Eq. \ref{g1}.
   The grey open symbols are a simple Drude formula, with a frequency independent scattering
   which is proportional to the temperature. }
   \label{stage1_g}
\end{figure}

\begin{figure}[ht]
   \centerline{\includegraphics[width=8cm,clip=true,angle=-90]{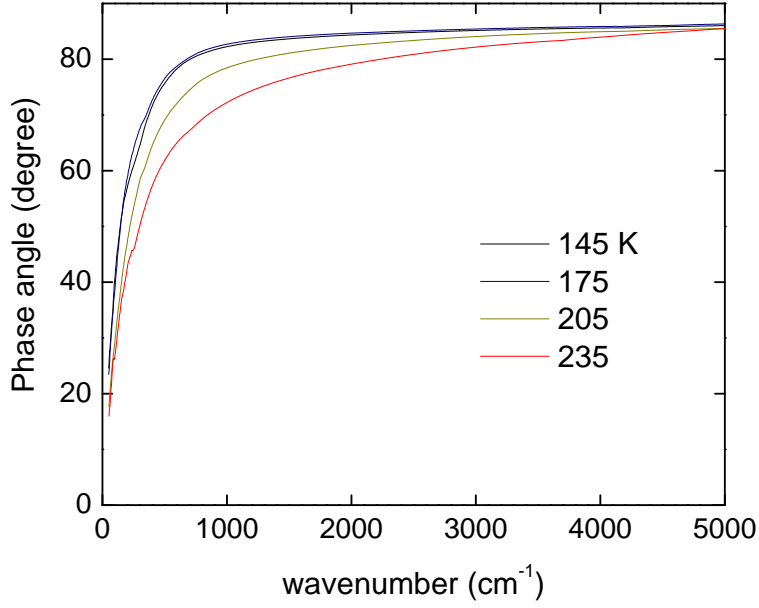}}
   \caption{Phase of $g(\omega,T)$ shown in Fig. \ref{stage1_g}}
   \label{stage1_g_phase}
\end{figure}

We have applied Eq. \ref{g1} to the full set of spectra in the
normal state (140-260 K in 1 K steps). The output for the
$g(\omega,T)$ is shown in Fig. \ref{stage1_g}. We see, that for a
broad range of values of $\hbar\omega/k_BT$ the scaling-function
is to a very good approximation given by the expression
\begin{equation}
g(\omega,T)=\frac{g(0)}{1+iA\hbar\omega/k_BT}
\end{equation}
where $A=0.83$ and $g(0)=1.67\cdot 10^{6} K\Omega^{-1}cm^{-1}$. This
is also borne out by the frequency dependent phase of this function,
displayed in Fig. \ref{stage1_g_phase}, which asymptotically
approaches 90 degree for $\omega/T\rightarrow\infty$. The curves for
the different temperatures now show a good scaling collapse, also
for $\hbar\omega/k_BT > 1.5$, where in the analysis of Ref.
\cite{marel03} the curves started to separate. Note, that the
imaginary part of $g(\omega,T)$ has more scatter than the real part,
which is due to the fact that for low frequencies the reflectivity
spectra depend in in leading order only on $\sigma_1$. This is the
so-called Hagen-Rubens limit, where for a frequency independent
$\sigma_1(\omega)$ the optical reflectivity has a $\omega^{0.5}$
departure from 1.

\begin{figure}[ht]
   \centerline{\includegraphics[width=8cm,clip=true,angle=-90]{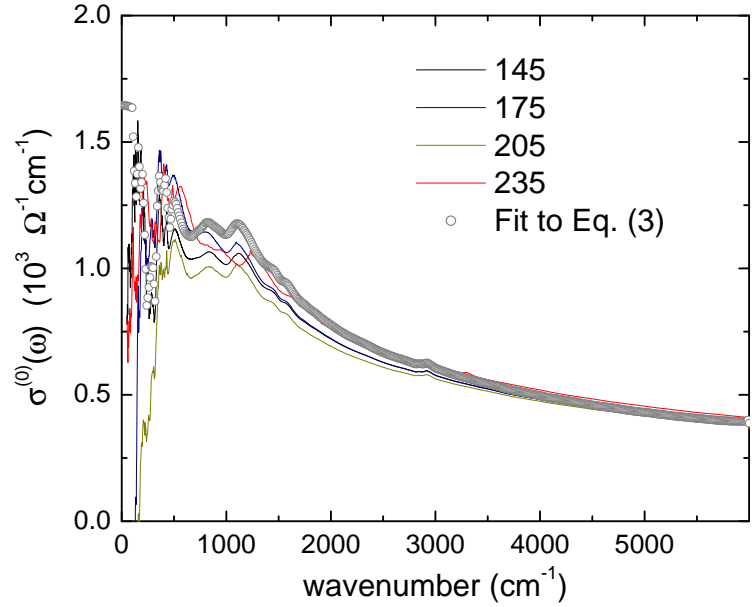}}
   \caption{Regular contribution to the optical conductivity $\sigma^{(0)}(\omega,T)$.
   Solid curves: Calculated by subtracting $Tg(\omega,T)$ shown in Fig.
   \ref{stage1_g} from the experimental spectra (see Fig. \ref{FigSig2223}). Grey circles:
   calculated by fitting the optical spectra at all temperatures between 140 K and 280 K
   from the experimental data using Eq. \ref{decomposition}. }
   \label{sigma(0)}
\end{figure}

\begin{figure}[ht]
   \centerline{\includegraphics[width=8cm,clip=true,angle=0]{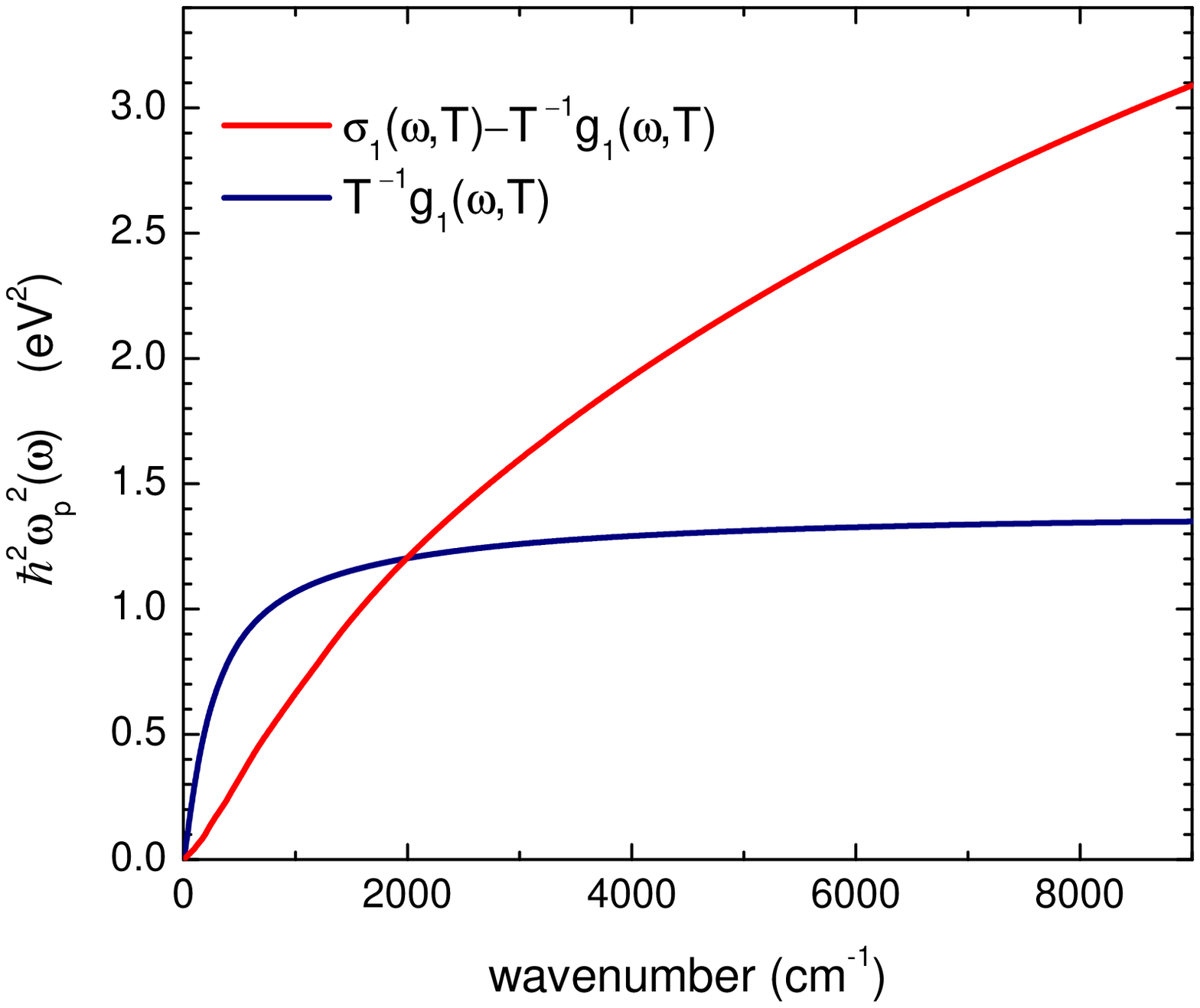}}
   \caption{Comparison of the spectral weight function at 235 K of the regular
   term (red curve)$\sigma^{(0)}(\omega,T)$ and $Tg(\omega,T)$}
   \label{stage1_sw}
\end{figure}
\begin{figure}[ht]
   \centerline{\includegraphics[width=8cm,clip=true,angle=-90]{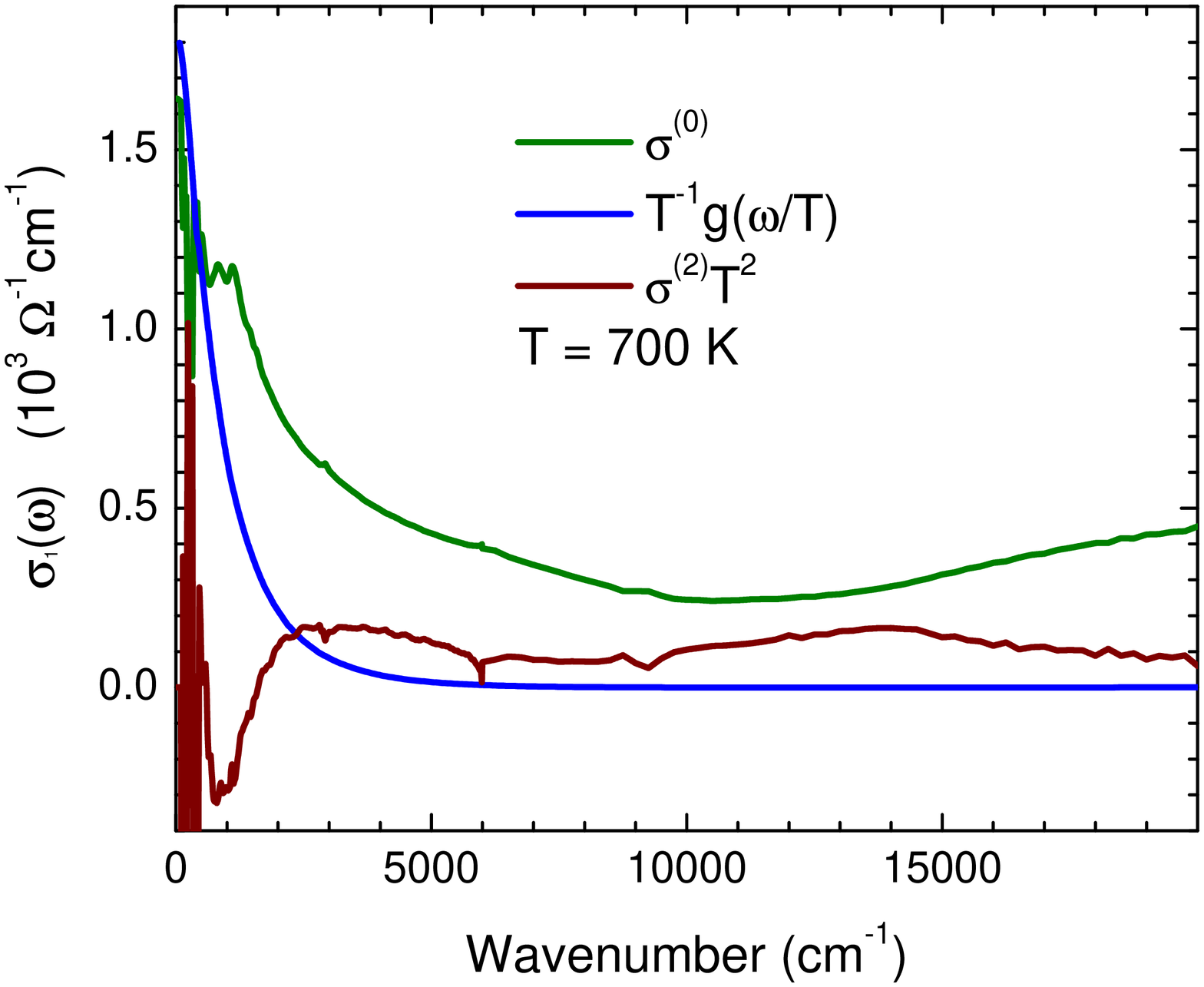}}
   \caption{Decomposition of the optical conductivity presented in Eq. \ref{decomposition}
   extrapolated to a high temperature ($T = 700$ K) in order to see all
   three contributions on the same scale.}
   \label{stage3combi}
\end{figure}

The good collapse of our scaling function for large
$\hbar\omega/k_BT > 1.5$ is clearly due to the fact, that in our
analysis we have removed a part of the optical conductivity
spectrum which is essentially independent of the temperature. This
background, which is obtained by subtracting $Tg(\omega,T)$ from
the experimental conductivity curves for the temperatures
indicated, $\sigma^{(0)}$, is displayed in Fig. \ref{sigma(0)}. If
the decomposition in a T-independent part and an $g(\omega/T)$
term would be perfect, these curves would be exactly on top of
each other. In fact, we see that this works quite well, but a few
significant changes remain. If we make instead the full
decomposition using Eq. \ref{decomposition} as discussed in
section\ref{stage3}, we obtain an almost identical result for
$g(\omega/T)$ and $\sigma^{(0)}$. These results are presented in
in Figs. \ref{sigma(0)} and \ref{stage3combi}.

The decomposition of the infrared spectra of the cuprates in a
Drude-peak and a so-called mid-infrared band has a long history,
dating back to the beginning of the high T$_c$ era. It has been
pointed out by Tanner and collaborators\cite{quijada99}, that if
one splits up the optical spectra this way, one finds that about
one quarter of the spectral weight below 1 eV resides in the Drude
peak. The remainder is in the temperature independent background.
Although our analysis is different in detail, the experimental
data do lead us to make an even stronger conclusion: If we try to
fit the data to a function of the general form $Tg(\omega/T)$ plus
a constant background, the result returned for $Tg(\omega/T)$ is a
simple Drude function with a T-linear scattering rate. That about
one third of the spectral weight or less resides in the Drude
peak, is borne out by Fig. \ref{stage1_sw}, where we display for
the two components the spectral weight function at T=245 K
\begin{equation}
\hbar^2\omega_p^2(\omega)=8\int_0^{\omega}Re\sigma(\omega)d\omega.
\end{equation}
\begin{figure}[ht]
   \centerline{\includegraphics[width=8cm,clip=true,angle=0]{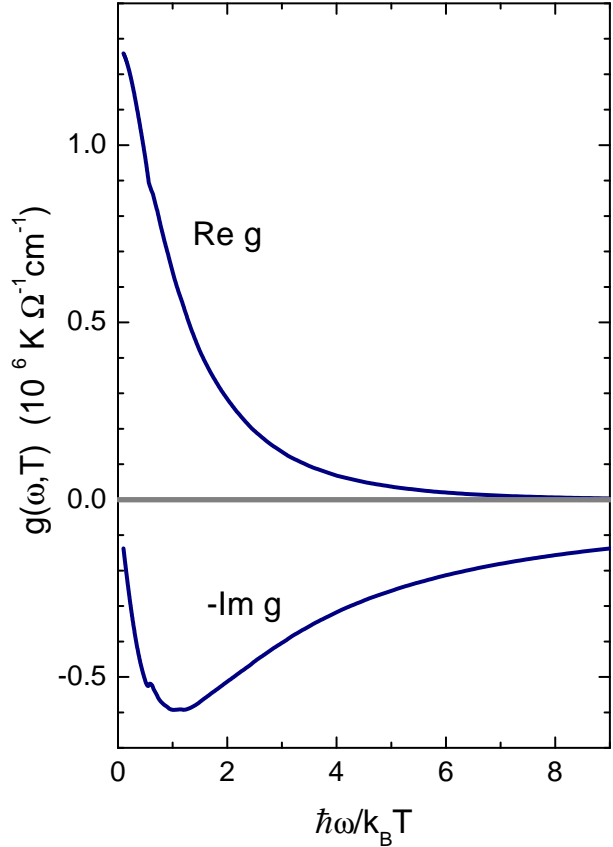}}
   \caption{Real and imaginary parts of the scaling function $g(\omega,T)$ of Bi2223,
   from a fit to the optical spectra at all temperatures between 140 K and 280 K from the experimental data using Eq. \ref{decomposition}.}
   \label{stage3_g}
\end{figure} 
%
%

%
\begin{itemize}

\item Our result indicates that the constant value of the phase
above 1000 cm$^{-1}$ is not a manifestation at high frequencies of
the same $g(\omega/T)$ function which dominates the low frequency
optical conductivity of the normal state.

\item If the $g(\omega/T)$ scaling collapse has something to do
with quantum criticality, then the constant phase above 1000
cm$^{-1}$ has a different origin, and vice versa.

\item {\em If} the decomposition of Eq. \ref{decomposition} would
be justified, our result shows that the scattering rate of
$g(\omega/T)$ has a linear dependence on the temperature, while it
is a constant as a function of frequency. There are several ways
of looking at this. Within a Fermi-liquid type approach this
appears to be an inconsistent result, because any reasonable model
which provides a temperature dependent scattering would have to
produce a similar energy dependent scattering, and vice versa.
This observation therefor appears to justify the assumption that
the decomposition of Eq. \ref{decomposition} is in fact not
justified in this case, and instead the entire optical
conductivity below 1 eV (or at least most of it) should be
interpreted as the high frequency response of the free charge
carriers.

\item Our result does not {\em imply} that the optical spectra of
the cuprates should be regarded as a sum of two independent terms,
one of which is essentially a constant of temperature and the
other a Drude term, it {\em assumes} such a decomposition. In
fact, the data in Figs. \ref{stage1_g} and \ref{sigma(0)} do not
show a perfect collapse of the curves at different temperatures,
and this already conveys a warning that Eq. \ref{decomposition}
may not be justified on a fundamental level.

\end{itemize}
The different terms in Eq.\ref{decomposition} are necessary to
provide a continuation of $g(\omega/T)$ for large $\omega/T$.
However, it seems overwhelmingly natural to interpret the entire
spectrum below 1 eV as the response of the free charge carriers.
In this view, the mid-infrared peak corresponds to the incoherent
part of the optical response. Whereas the Drude peak results from
the collective response to an electric field of the
quasi-particles, the incoherent part results from the fact that
those quasi-particles are formed out of ordinary electrons due to
the coupling to a bosonic spectrum, $P(\omega)$. The Kubo formula
for the optical conductivity is
\begin{eqnarray}\label{kubo1}
T\sigma(\omega,T) = \int dx
\frac{\omega_p^2}{4\pi
i\omega/T}\frac{(e^{x}+1)^{-1}-(e^{x+\omega/T}+1)^{-1}}{\omega/T-s^{\star}(\epsilon)+
s(\omega+\epsilon)}
\end{eqnarray}
where $s(\omega)$ is the self energy devided by the temperature,
$s(\epsilon)=\Sigma(\omega)/T $, and $\Sigma(\omega)$ follows from
perturbation theory of the electron-boson coupling\cite{norman06}
\begin{eqnarray}\label{kubo2}
s(\omega)\equiv \int dx\int dy
\frac{(e^y-1)^{-1}+(e^x+1)^{-1}}{\epsilon/T+y-x+i0^+} P(y,T)
\end{eqnarray}
In the Fermi and Bose functions of the above expressions
energy/temperature ratios have been substituted with the
dimensionless integration variables $x$ and $y$ respectively.
Varma {\em et al.}\cite{varma89,abrahams96} have proposed a
bosonic spectrum with the remarkable property, that it contains no
low energy scale other than the temperature itself
\begin{eqnarray}\label{bosons}
\nonumber P(y,T) =& Cy  & \mbox{        } (y<1)  \\
\nonumber      =& C   & \mbox{        } (1 < y < \omega_c/T) \\
               =& 0   & \mbox{        } (y > \omega_c/T)
\end{eqnarray}
The high frequency cutoff $\omega_c$ causes the quasi-particle
residue at the Fermi surface to vanish logarithmically at low
temperature and frequency\cite{varma89,abrahams96}. Without
$\omega_c$, the function $P(y)$ would have no explicit dependence
on the temperature, and it is clear from the structure of Eqs.
\ref{kubo1} and \ref{kubo2}, that the explicit temperature
dependence would disappear from $T\sigma(\omega,T)$ as well. In
other words, the optical conductivity would be of the form
$\sigma(\omega,T)=T^{-1}g(\omega/T)$ as discussed in this article.
In this manuscript we have shown that if we assume that the
optical response is of the form
$\sigma(\omega,T)=T^{-1}g(\omega/T)$ plus a T-independent
background spectrum, the function $g(\omega/T)$ is characterized
by a scattering rate which varies linearly as a function of
temperature but is a constant as a function of frequency. At first
glance our data therefor appear to disagree with the Ansatz of Eq.
\ref{bosons}. However, at present our analysis can not take into
account the possible influence of a high energy cutoff $\omega_c$.
At this stage, based on the arguments presented in this manuscript
we can not rule out nor confirm the validity of Eq. \ref{bosons}
for the optical spectra of the optimally doped cuprates. In
particular the extent to which $\omega_c$ can influence the
optical spectra at much lower frequencies needs to be
investigated.

In a recent paper Norman and Chubukov demonstrated, that the
observed phase angle and frequency exponent of the modulus of the
conductivity can arise from scattering by a continuum of bosons
extending at least to 300 meV\cite{norman06}, a value for which
neither lattice vibrations nor some fluctuating order parameter is
sensible. They pointed out, that this high frequency cut-off is
evident as a peak in Im$1/\sigma(\omega)-4\pi\omega/\omega_p^{2}$.
Since the factor $\omega_p^{2}$ follows from the f-sum rule, such
a graph can be generated from the experimental data without
further assumptions. A more radical view on the nature of the
excitations studied with optical spectroscopy has been advocated
by P.W. Anderson, who points out that, due to the large on-site
repulsion, the excitations are not true quasiparticles in the
Landau sense\cite{anderson06}. When $\omega/T > 1$, the
excitations created by absorbing a photon are scattered before
they can recohere into the original quasiparticle. In this regime
the conductivity of a sample with a hole doping $x$ should be a
power law\cite{anderson06},

\begin{eqnarray}
\sigma(\omega)\propto (i\omega)^{-1+3p}
\mbox{   where   } p=(1-x)^2/8
\end{eqnarray}
Indeed the experimentally observed exponent is very close to what
one expects based on the known value of the doping, $x$. An
important point which still requires further theoretical and
experimental attention concerns the way in which the $\omega>T$ is
connected to the low frequency range, $\omega<T$ range. The
detailed behavior of this cross-over promises to be an excellent
way to compare, and perhaps distinguish, the predictions based on
different theoretical models.

\section{Conclusions}
We have analyzed in detail the possibility that the optical
conductivity of the cuprates at optimal doping is described by a
universal scaling formula
$\sigma(\omega,T)=T^{-1}g(\omega/T)+\sigma^{(0)}(\omega)$. We have
introduced a differential-integral transformation which allows to
extract both $g(\omega/T)$ and $\sigma^{(0)}(\omega)$ from the
experimental data, provided that these have been measured with
sufficient accuracy and in sufficient dense set of temperatures.
We have applied this analysis to recent high precision data of
Bi2223. From our analysis we conclude, that $g(\omega/T)$
corresponds to a simple Drude formula with a scattering rate which
depends linearly on the temperature, and which has no frequency
dependence. The 'regular' contribution $\sigma^{(0)}(\omega)$ is
almost a constant as a function of frequency and temperature, and
contributes two thirds of the spectral weight below 1 eV. The
phase of the total, experimental, optical conductivity is a
constant as a function of frequency above ~0.1 eV. This constant
value is a result of the contributions from both
$T^{-1}g(\omega/T)$ and $\sigma^{(0)}(\omega)$, hence it is not a
continuation of the scaling function $g(\omega/T)$ to high
frequencies. Consequently, if the $g(\omega/T)$ scaling collapse
has something to do with quantum criticality, then the constant
phase above 1000 cm$^{-1}$ has probably a different origin, and
vice versa.

\section{Acknowledgements}
We gratefully acknowledge useful discussions about this subject
with C. M. Varma, E. Abrahams, J. Zaanen, M. R. Norman, A. V.
Chubukov, F. Marsiglio, A. J. Millis, P. W. Anderson and N.
Bontemps. This work was supported by the Swiss National Science
Foundation through the National Center of Competence in Research
"Materials with Novel Electronic Properties-MaNEP".

\end{document}